\documentclass[floatfix,prb,groupedaddress,twocolumn]{revtex4}

\usepackage{graphicx}

\begin{document}

\title{Single-ensemble nonequilibrium path-sampling
estimates of free energy differences}
\author{F.\ Marty Ytreberg}
\author{Daniel M.\ Zuckerman}
\affiliation{Center for Computational Biology and Bioinformatics,
  School of Medicine}
\affiliation{ Dept.\ of Environmental and Occupational Health,
  Graduate School of Public Health,
  University of Pittsburgh, 200 Lothrop St., Pittsburgh, PA 15261}
\date{\today}

\begin{abstract}
We introduce a straightforward, single-ensemble, path sampling
approach to calculate free energy differences based on
Jarzynski's relation. For a two-dimensional ``toy'' test
system, the new (minimally optimized) method performs roughly one
hundred times faster than either optimized ``traditional'' Jarzynski
calculations or conventional thermodynamic integration.
The simplicity of the underlying formalism
suggests the approach will find broad applicability in molecular
systems.
\end{abstract}


\maketitle

The estimation of free energy differences $\Delta F$
in molecular systems
\cite{zuckerman-prl,bustamante-sci,sun,jarzynski,oostenbrink,
mordasini,hummer,schulten-2003,shirts-jcp}
is important for a wide variety of applications
including virtual screening for drug design,
determination of the solubility of
small molecules, and binding affinities of ligands to
proteins \cite{leach-book,burgers}.
Jarzynski recently introduced a general non-equilibrium
approach to computing $\Delta F$ \cite{jarzynski,crooks-pre},
but the technique never has
been shown superior to more traditional equilibrium
calculations (e.g.\ Refs.\ \cite{hummer,mordasini}). Here, we introduce
a potential route for dramatically faster non-equilibrium $\Delta F$
calculations.

Many previous workers have attempted to improve non-equilibrium
$\Delta F$ estimates. Hummer studied the optimization of
non-equilibrium simulation \cite{hummer}, and Jarzynski
introduced ``targeted free energy perturbation''
to improve configurational sampling \cite{jarzynski-targeted}.
Improvement of configurational
sampling in $\Delta F$ calculations has also been
the focus of studies by McCammon and collaborators \cite{mordasini},
Karplus and collaborators \cite{karplus-cpl} and 
van Gunsteren and collaborators \cite{oostenbrink}.
Schulten and collaborators used
Jarzynski's approach for steered molecular dynamics
simulations \cite{schulten-2003}.
Ytreberg and Zuckerman \cite{ytreberg-extrap}, and
Zuckerman and Woolf 
\cite{zuckerman-prl,zuckerman-cpl} have developed methods for
more efficient use of non-equilibrium data for $\Delta F$
calculation.

In an important advance of direct relevance to the present
report, Sun suggested the use of a path sampling
approach to evaluate $\Delta F$ via Jarzynski's relation,
with a formalism that essentially entails
thermodynamic integration
in (inverse) temperature space \cite{sun}.
Sun reported impressive efficiency gains. However,
multiple path sampling ensembles
were required even for simple systems.

The approach outlined below builds on
several sources.
Jarzynski defined the non-equilibrium approach \cite{jarzynski},
and Pratt introduced the seminal concept of sampling dynamic paths
with equilibrium tools \cite{pratt}. Chandler and collaborators
supplied Monte Carlo path sampling moves for effective
implementation of the Pratt approach
\cite{tps-review,dellago-rate,crooks-tps},
and Sun suggested that path sampling
ensembles could be used to evaluate the Jarzynski relation
\cite{sun}. Finally, Zuckerman and Woolf employed a
direct formalism for path-based estimates
of arbitrary quantities, which is key to our single-ensemble
protocol \cite{zuckerman-tps}.

In outline, this report first sketches Jarzynski's relation and
shows how it can be re-written using importance sampling of paths.
The path sampling procedure used in our method is then described.
Finally, we present our results and a discussion.

Following the usual formalism to define
the $\Delta F$ calculation, we consider two systems
or distinct states that are defined by
Hamiltonians $H_0(\vec{x})$ and $H_1(\vec{x})$, where $\vec{x}$
is a set of configurational coordinates. By introducing
a parameter $\lambda$, a hybrid Hamiltonian can be constructed,
e.g.,
$H(\lambda;\vec{x})=H_0(\vec{x})+
\lambda \bigl[ H_1(\vec{x})-H_0(\vec{x})\bigr]$.
Jarzynski showed that arbitrarily rapid,
non-equilibrium switches from
$\lambda=0$ to $\lambda=1$ can be used to calculate the
{\it equilibrium}
free energy difference $\Delta F=\Delta F_{\lambda=0 \rightarrow 1}$.
To this end, one considers switching trajectories that combine
increments in $\lambda$
with ``traditional'' dynamics (such as Monte Carlo or
Langevin dynamics) in $\vec{x}$-space at fixed $\lambda$
values. Thus, a trajectory with $n$ $\lambda$-steps is given by
\begin{eqnarray}
  {\bf Z}_n = \Bigl\{
    (\lambda_0=0,\vec{x}_0),(\lambda_1,\vec{x}_0),
    (\lambda_1,\vec{x}_1),(\lambda_2,\vec{x}_1), \nonumber \\
    (\lambda_2,\vec{x}_2),...,
    (\lambda_{n-1},\vec{x}_{n-1}),(\lambda_n=1,\vec{x}_{n-1})
    \Bigr\},
  \label{eq-traj}
\end{eqnarray}
where it should be noted that increments (steps) from
$\lambda_i$ to $\lambda_{i+1}$
are performed at a fixed conformation $\vec{x}_i$, and
the initial $\vec{x}_0$ is drawn from
the $H_0$ distribution. For simplicity we have assumed
only a single dynamics step at each fixed $\lambda_i$,
from $\vec{x}_{i-1}$ to $\vec{x}_i$,
is performed, but multiple steps can be
performed within the Jarzynski formalism.

Finally, the work performed on the system
during a switching trajectory is
\begin{eqnarray}
  W({\bf Z}_n)=\sum_{i=0}^{n-1}\Bigl[
    H(\lambda_{i+1};\vec{x}_i)-H(\lambda_i;\vec{x}_i)
    \Bigr],
  \label{eq-work}
\end{eqnarray}
and transcribing the Jarzynski relation \cite{jarzynski}
into path language, the free energy difference can be written
as \cite{sun}
\begin{eqnarray}
  {\rm e}^{-\beta \Delta F}=
  {\int d{\bf Z}_n \; Q({\bf Z}_n) \; {\rm e}^{-\beta W({\bf Z}_n)}}
  \bigg/
  {\int d{\bf Z}_n \; Q({\bf Z}_n)},
      \label{eq-jarz}
\end{eqnarray}
where $\beta=1/k_BT$, $d{\bf Z}_n$ denotes integration over
all possible
trajectories, and $Q({\bf Z}_n)$ is proportional
to the probability of occurrence of trajectory ${\bf Z}_n$.
$Q({\bf Z}_n)$ depends on the dynamics employed
and will be specified
below for the overdamped Langevin case.

In ``standard'' non-equilibrium simulation, the
integral (\ref{eq-jarz})
need never be considered since trajectories --- and the
associated work values --- are automatically generated
with the proper frequency (i.e., proportional to
$Q({\bf Z}_n)$).
In this case, the Jarzynski relation provides an
estimate for $\Delta F$ for a set of work
values $\{W_1,W_2,...,W_N\}$ given by \cite{jarzynski}
\begin{eqnarray}
  \Delta F \doteq \Delta F_{\rm Jarz} \equiv
  -\frac{1}{\beta}\ln \left[
    \frac{1}{N}\sum_{i=1}^N {\rm e}^{-\beta W_i}
    \right],
  \label{eq-jarzest}
\end{eqnarray}
where the ``$\doteq$'' denotes a computational estimate.
Since the relationships in Eqs.\ (\ref{eq-jarz})
and (\ref{eq-jarzest}) are valid
for an arbitrary number $n$ of $\lambda$-steps,
switches may be performed very rapidly. The apparent
advantage of these ``fast-growth'' (small $n$) calculations
is that very little computational time is spent generating
trajectories, and thus $\Delta F_{\rm Jarz}$
can be generated with very little CPU time.
However, in practice,
unless there is sufficient overlap between the states
described by $H_0(\vec{x})$ and $H_1(\vec{x})$,
$\Delta F_{\rm Jarz}$ will be biased,
often by many $k_BT$
\cite{oostenbrink,jarzynski-targeted,mordasini}.
This bias is due to the nonlinear
nature of Eq.\ (\ref{eq-jarzest}) where the smallest,
and thus {\it rarest}, work values dominate the average. 
Additionally, CPU time must be invested in generating
the equilibrium distribution for $H_0$.

This study uses importance sampling of switching
trajectories to sample
dominant but rare work values more frequently, without
the need to sample the $H_0$ equilibrium distribution.
We combine the sampling strategy of Sun \cite{sun} with the simple
formalism used by Zuckerman
and Woolf \cite{zuckerman-tps}, as we consider
an alternative distribution of switching trajectories
$D({\bf Z}_n)$.
Then, with no loss of generality, Eq.\ (\ref{eq-jarz})
can be written as
\begin{eqnarray}
  {\rm e}^{-\beta \Delta F}=
  \frac
      {\int d{\bf Z}_n \; D({\bf Z}_n)\Bigl[
	  Q({\bf Z}_n)/D({\bf Z}_n)\Bigr]
	\; {\rm e}^{-\beta W({\bf Z}_n)}}
      {\int d{\bf Z}_n \; D({\bf Z}_n)\Bigl[
	  Q({\bf Z}_n)/D({\bf Z}_n)\Bigr]} \nonumber \\
	\doteq
	\frac
	    {\sum^D Q({\bf Z}_n) \; {\rm e}^{-\beta W({\bf Z}_n)}/D({\bf Z}_n)}
	    {\sum^D Q({\bf Z}_n)/D({\bf Z}_n)}
	    \label{eq-isgen}
\end{eqnarray}
where the only condition is that $D({\bf Z}_n) \neq 0$ anywhere.
The shorthand $\sum^D$ indicates a sum over trajectories
generated according to $D({\bf Z}_n)$.

Since the fundamental idea behind the importance sampling
in Eq.\ (\ref{eq-isgen}) is to generate trajectories --- 
and hence work values --- according to
$D({\bf Z}_n)$, the choice $D$ is critical.
We choose $D({\bf Z}_n)$ to favor trajectories with
important work values, namely,
\begin{eqnarray}
  D({\bf Z}_n) =
  Q({\bf Z}_n){\rm e}^{-\frac{1}{2}\beta W({\bf Z}_n)}.
  \label{eq-rho}
\end{eqnarray}
As will be seen below in Eq.\ (\ref{eq-isest}), this choice
appears to balance convergence difficulties between the
numerator and denominator of Eq.\ (\ref{eq-isgen}).
We note that Sun also employed the distribution
(\ref{eq-rho}) as one
among several used for an indirect calculation of
$\Delta F$ \cite{sun}.
While it is not obvious that the choice (\ref{eq-rho})
is optimal in general, other forms
for $D({\bf Z}_n)$ have been tested by the authors and provided
no improvement. By comparison with (\ref{eq-jarz}), the $\beta/2$
in (\ref{eq-rho}) embodies double the temperature.

Combining Eqs.\ (\ref{eq-isgen}) and (\ref{eq-rho}), the
free energy estimate for our single-ensemble
path sampling (SEPS) method is given by the new relation
\begin{eqnarray}
  \Delta F \doteq \Delta F_{\rm seps} \equiv
  -\frac{1}{\beta}\ln \biggl[
    {{\sum}^D {\rm e}^{-\frac{1}{2}\beta W}}\bigg/\;
	 {{\sum}^D {\rm e}^{+\frac{1}{2}\beta W}}
	 \biggr].
  \label{eq-isest}
\end{eqnarray}

We now specify $Q({\bf Z}_n)$ from Eq.\ (\ref{eq-isest})
which is required for the path sampling performed below.
We assume overdamped Langevin (Brownian) dynamics
is used at fixed $\lambda$ values. Single-step distributions for
$\Delta \vec{x}_i=\vec{x}_i-\vec{x}_{i-1}$ are thus Gaussian,
with a variance given by $\sigma^2=2\Delta t/m\gamma\beta$,
where $m$ is the mass of the particle
and $\gamma$ is the friction coefficient of the
medium (e.g.\ Ref.\ \cite{zuckerman-tps}).
Combining the Brownian distributions with that
for $\lambda=0$  
leads to the full trajectory weight
\begin{eqnarray}
  Q({\bf Z}_n)=
  {\rm e}^{-\beta H_0(\vec{x}_0)}
  \prod_{i=1}^{n-1}
  \frac
      {{\rm e}^{-| \Delta \vec{x}_i-\Delta \vec{x}_i^{\rm det}|^2 /2\sigma^2}}
      {(2\pi\sigma^2)^{d/2}}
    \label{eq-bdweight}
\end{eqnarray}
where $\Delta \vec{x}_i^{\rm det}=-\vec{\nabla}_x
H(\lambda_i;\vec{x}_{i-1})\Delta t/m\gamma$ is
proportional to the force and time step,
and $d$ is the dimensionality of the conformational space
$\vec{x}$. Note that if deterministic dynamics (e.g., Verlet)
is used then $Q({\bf Z}_n)={\rm e}^{-\beta H_0(\vec{x}_0)}$.

To calculate the free energy estimate $\Delta F_{\rm seps}$
in Eq.\ (\ref{eq-isest}),
switching trajectories must be generated according to
$D({\bf Z}_n)$. This is readily accomplished using the 
path sampling approach proposed by Pratt \cite{pratt},
where entire trajectories (paths) are
generated and then accepted or rejected based upon a suitable 
Monte Carlo criteria. Trial moves in path space are generated
following Chandler and coworkers \cite{tps-review,dellago-rate}.

Putting the pieces together, we estimate $\Delta F$
by sampling trajectories
according to $D({\bf Z}_n)$ in Eq.\ (\ref{eq-rho}) using the
following steps (c.f.\ Ref.\ \cite{sun}):
(i) Generate an arbitrary initial reference trajectory
by switching the
system from $\lambda=0 \rightarrow 1$. Calculate the
work $W$ done on the system during the switch.
(ii) Pick a random $\lambda$ value along the reference
trajectory and make a random phase-space displacement.
For Brownian dynamics
this corresponds to a random shift in position.
Generate a trial trajectory by ``shooting'' forward
(increment $\lambda$) and backward (decrement $\lambda$).
Calculate the trial work done on the system $W'$.
(iii) Accept this new trajectory according to the Metropolis
criterion: $\min\bigl[
  1,{Q^{\prime}{\rm e}^{-\frac{1}{2}\beta W'}}\big/
  {Q{\rm e}^{-\frac{1}{2}\beta W}}
  \bigr],$
with $Q$ from Eq.\ (\ref{eq-bdweight}).
(iv) If accepted, the trial trajectory becomes the current
reference trajectory. If rejected, the current reference
trajectory remains unchanged. Whether accepted or rejected,
the current reference trajectory is then used in
Eq.\ (\ref{eq-isest}). Repeat from step (ii).

It should be noted that to obtain good sampling, as in any
Monte Carlo simulation, equilibrium must
be attained before averages are calculated.
Using the path sampling procedure above,
we accomplish this by checking the running average
work every 20 accepted trajectories
for convergence within 0.01 $k_BT$.

\begin{figure}
  \includegraphics[scale=0.34]{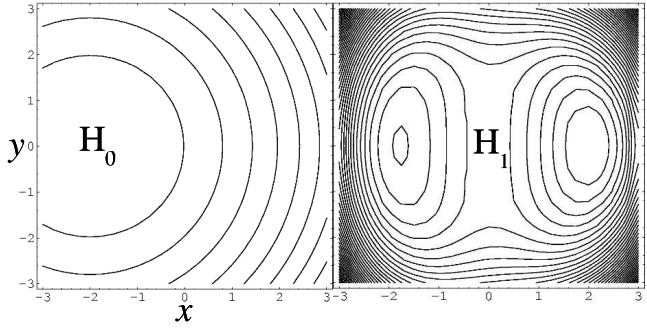}
  \caption{
    Contour plots of the test system
    $H_0(x,y)$ and $H_1(x,y)$ given by
    Eq.\ (\ref{eq-pot}) where each contour
    represents
    an energy change of 4.0 $k_BT$.
    This problem is expected
    to be difficult due to the large barrier
    and the asymmetric double-well in $H_1$.
    \label{fig-pot}
  }
\end{figure}

As a test problem, we consider a two-dimensional system
(Fig.\ \ref{fig-pot}) switched from a single well to a
double well:
\begin{eqnarray}
  H_0(x,y)=(x+2)^2+y^2, \nonumber\\
  H_1(x,y)=\frac{1}{10}
  \Bigl\{
  ((x-1)^2-y^2)^2+10(x^2-5)^2+\nonumber\\
  (x+y)^4+(x-y)^4
  \Bigr\}.
  \label{eq-pot}
\end{eqnarray}
Figure \ref{fig-pot} clearly demonstrates why
estimating $\Delta F$ for this system is expected to be
difficult: the significant barrier in $H_1$
will prevent sufficient configurational
sampling of the dominant minimum at $H_1(2,0)$
for short trajectories. Thus
ordinary fast-growth Jarzynski estimates will substantially
overestimate $\Delta F$. Similarly, equilibrium approaches
like thermodynamic integration (TI) will require long
simulation times to surmount the barrier.

For this system, the free energy difference was estimated
using the Jarzynski method given by $\Delta F_{\rm Jarz}$
in Eq.\ (\ref{eq-jarzest}),
by the SEPS method given by $\Delta F_{\rm seps}$
in Eq.\ (\ref{eq-isest}), as well as by conventional TI.
Trajectories for all estimates were generated using
Brownian dynamics with parameters $\beta=\gamma=m=1$,
and $\Delta t=0.001$.

For $\Delta F_{\rm Jarz}$, to generate uncorrelated
initial configurations $\vec{x}_0$, the system was run
at $\lambda=0$ for $N_{\rm eq}$ steps between switching
trajectories. For $H_0$ given by (\ref{eq-pot}), it was
determined that $N_{\rm eq}=10,000$,
and that smaller values of
$N_{\rm eq}$ introduce bias in $\Delta F_{\rm Jarz}$.
Given $N_{\rm eq}$, moreover, we {\it optimized
$\Delta F_{\rm Jarz}$
by varying the number $n$ of $\lambda$-steps}
in Eqs.\ (\ref{eq-traj}) and (\ref{eq-work}). We found
that $n=100,000$ was most efficient.

Trajectories for $\Delta F_{\rm seps}$ were generated
as described above.
Specifically, perturbations to the selected
state $\vec{x}_i$ of the reference trajectory
(step (ii) above) were chosen
from a Gaussian distribution of width $50.0\sigma$,
giving an acceptance ratio of $1-2\%$.
Smaller perturbations were also highly successful.
The SEPS procedure is {\it not optimized} in the sense
that only a simple type of trial move (termed
``shooting'' \cite{tps-review}) was employed,
and we used strict path-equilibration criteria.
Optimization methods are currently
under investigation by the authors.

For comparison to an equilibrium approach, ``textbook''
thermodynamic integration (TI) simulations
\cite{frenkel-book} were performed with
identical Brownian parameters and 10 $\lambda$-steps,
with 25\% of data discarded for equilibration. 
This well-known approach is described in many sources
(e.g., Ref.\ \cite{karplus-jcp}) and is not detailed here.
Since {\it no optimization was performed},
we refer to this method as ``conventional TI.''

\begin{figure}
  \includegraphics[scale=0.34]{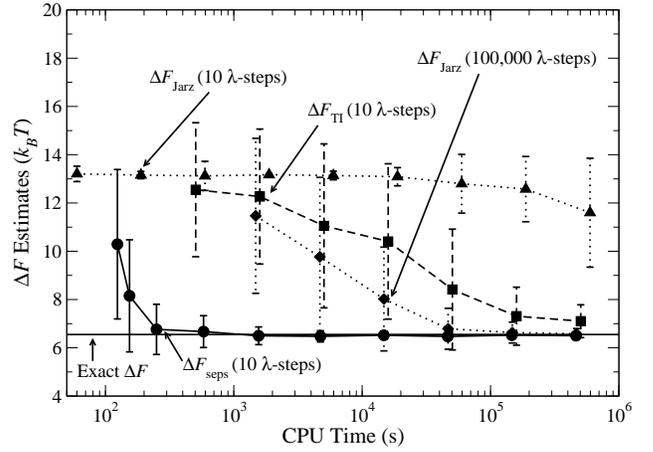}
  \caption{
    Comparison between free energy estimates
    from the Jarzynski method,
    conventional Thermodynamic integration (TI),
    and our single-ensemble path sampling (SEPS)
    method. The circles show the results of the SEPS
    method using 10 $\lambda$-steps. The results
    of the Jarzynski method for a very short trajectory
    (10 $\lambda$-steps, squares) and the most efficient
    trajectory length (100,000 $\lambda$-steps, triangles)
    are also shown.
    TI estimates based on 10 $\lambda$ increments
    are shown as diamonds.
    The exact answer of $\Delta F=6.55 \, k_BT$
    is shown as a solid horizontal line.
    Each data point represents the mean
    estimate, with standard deviations
    given by the error bars, based on 
    100 independent estimates of $\Delta F$
    for each method.}
  \label{fig-results}
\end{figure}

To compare the efficiency of the SEPS approach
with other methods, in Fig.\ \ref{fig-results}
we plot $\Delta F$ estimates for the SEPS, 
Jarzynski, and TI methods as a function of
the total CPU time needed generate
the estimates. The circles show the results
of the SEPS method 
using 10 $\lambda$-steps. Also shown are
the results of the Jarzynski method using
a very short trajectory
(10 $\lambda$-steps, squares), and
the most efficient trajectory length
(100,000 $\lambda$-steps, triangles).
TI estimates based on 10 $\lambda$ increments
are shown as diamonds.
The solid horizontal line gives the exact answer
$\Delta F=6.55 \, k_BT$. The plot was
generated by calculating the mean (data points)
and standard deviations (error bars) from
100 independent estimates of $\Delta F$
for each method. The CPU time spent
equilibrating is included in the total CPU time
for all methods.

As expected, Fig.\ \ref{fig-results} shows that
for fast-growth work values
(10 $\lambda$-steps, squares),
the Jarzynski method incorrectly estimates the
free energy difference as
$\Delta F_{\rm Jarz} \approx 13 \, k_BT$. As the number
of $\lambda$-steps increases, the standard Jarzynski
trajectories begin to ``see'' the minimum at $H_1(2,0)$
and the correct $\Delta F$ is obtained.
Since the highest efficiency for the standard Jarzynski method
was obtained using 100,000 $\lambda$-steps (triangles),
we consider this curve to be the optimized Jarzynski method
for the test system.
The unoptimized, conventional TI calculations are of
comparable efficiency to the traditional Jarzynski estimates.

The SEPS method, by contrast, correctly estimates the
free energy quickly and accurately, even for
very short trajectories (10 $\lambda$-steps). One can
quantitatively compare estimates from each
method by noting that the estimate for
the SEPS method $\Delta F_{\rm seps}(t\approx 1500$ s)
is slightly more accurate than
$\Delta F_{\rm Jarz}(t\approx 150,000$ s)
and $\Delta F_{\rm TI}(t \approx 500,000$ s),
implying a roughly 100-fold speed-up of SEPS
over the other methods.

Compared to ``standard'' Jarzynski calculation, the SEPS
approach has several advantages:
(a) important, rare trajectories with small work values
are favored;
(b) no CPU time is spent acquiring an equilibrium ensemble
at $\lambda=0$; and
(c) path-sampling moves that are capable of surmounting
barriers may be used.
In other words, the SEPS approach focuses CPU time on the important
regions of ($\lambda ; \vec{x} $) space --- this also
contrasts with TI and other equilibrium approaches which
attempt to sample the full space. 

To summarize, we have described a rapid and straightforward
new method for estimating
free energy differences $\Delta F$, using
a single-ensemble path sampling (SEPS) approach.
We also have carefully quantified the numerical
efficiency of the approach.
Without extensive optimization,
the SEPS method generates $\Delta F$ estimates
over 100$\times$ more efficiently than ``standard'' 
Jarzynski and conventional thermodynamic integration
calculations, for the two-dimensional test system
considered here. Our approach relies on an extremely
simple importance
sampling formalism, and therefore appears to be readily
extendable to molecular systems. This extension --- 
which will require addressing issues of memory for trajectory
storage --- is
currently underway.
We will also optimize the SEPS approach via
alternative importance sampling distributions,
and path-sampling trial moves.

We would like to thank Arun Setty and Robert Swendsen
for insightful discussion.
Funding for this research was provided by
the Dept.\ of Environmental and Occupational
Health at the University of Pittsburgh, and
the National Institutes of Health (Grant T32 ES007318).

\bibliography{/home/marty/research/other/tex/bib/my}

\end{document}